\documentclass[aps,twocolumn,floatfix,footinbib,groupedaddress,superscriptaddress,showpacs]{revtex4-1}
\usepackage{graphicx}
\usepackage{mathrsfs}

\usepackage{amsmath}
\usepackage{amssymb}
\usepackage{latexsym} 
\usepackage{amsfonts} 
\usepackage{epsfig}
\usepackage{psfrag} 
\usepackage[latin1]{inputenc}
\usepackage[english]{babel}
\usepackage{graphicx}%
\usepackage{dcolumn}%
\usepackage{multirow}
\usepackage[table]{xcolor}
\usepackage{booktabs}
\usepackage{soul} 
\usepackage{hyperref}

\newcommand{\ket}[1] {\left| #1 \right\rangle}
\def\openone{\leavevmode\hbox{\small1\kern-4.2pt\normalsize1}}

\newcommand{\nn}{\nonumber\\}
\newcommand{\bea}{\begin{eqnarray}}
\newcommand{\ea}{\end{eqnarray}}

\hypersetup{
    colorlinks=true,       
    linkcolor=blue,          
    citecolor=blue,        
    urlcolor=blue        
}

\begin{document}

\title{Towards Outperforming Classical Algorithms with Analog Quantum Simulators}
\author{Sarah Mostame$^1$, Joonsuk Huh$^1$, Christoph Kreisbeck$^1$, Andrew J Kerman$^2$, Takatoshi Fujita$^1$, 
Alexander Eisfeld$^{\dagger \, 3}$, and Al\'an Aspuru-Guzik$^*\,$
}
\address{
Department of Chemistry and Chemical Biology, Harvard University, Cambridge, MA 02138, USA
\\
$^2$Lincoln Laboratory, Massachusetts Institute of Technology, Lexington, MA 02420, USA
\\
$^3$Max-Planck-Institut f\"ur Physik komplexer Systeme, N\"othnitzer Str. 38, 01187 Dresden, Germany
}
\begin{abstract}
With quantum computers being out of reach for now, quantum simulators are the alternative devices for
efficient and more exact simulation of problems that are challenging on conventional computers.
Quantum simulators are classified into analog and digital, with the possibility of constructing
``hybrid'' simulators by combining both techniques.
In this paper, we focus on analog quantum simulators of open quantum systems and address the limit that
they can beat classical computers. 
In particular, as an example, we discuss simulation of the chlorosome light-harvesting antenna from green 
sulfur bacteria with over 250 phonon modes coupled to each electronic state.
Furthermore, we propose physical setups that can be used to reproduce the quantum dynamics of a standard 
and multiple-mode Holstein model.
%
The proposed scheme is based on currently available technology of superconducting circuits consist of flux qubits and quantum oscillators.

\end{abstract}

\maketitle
\begin{figure}[t]
\begin{center}
\resizebox{\linewidth}{!}{\includegraphics{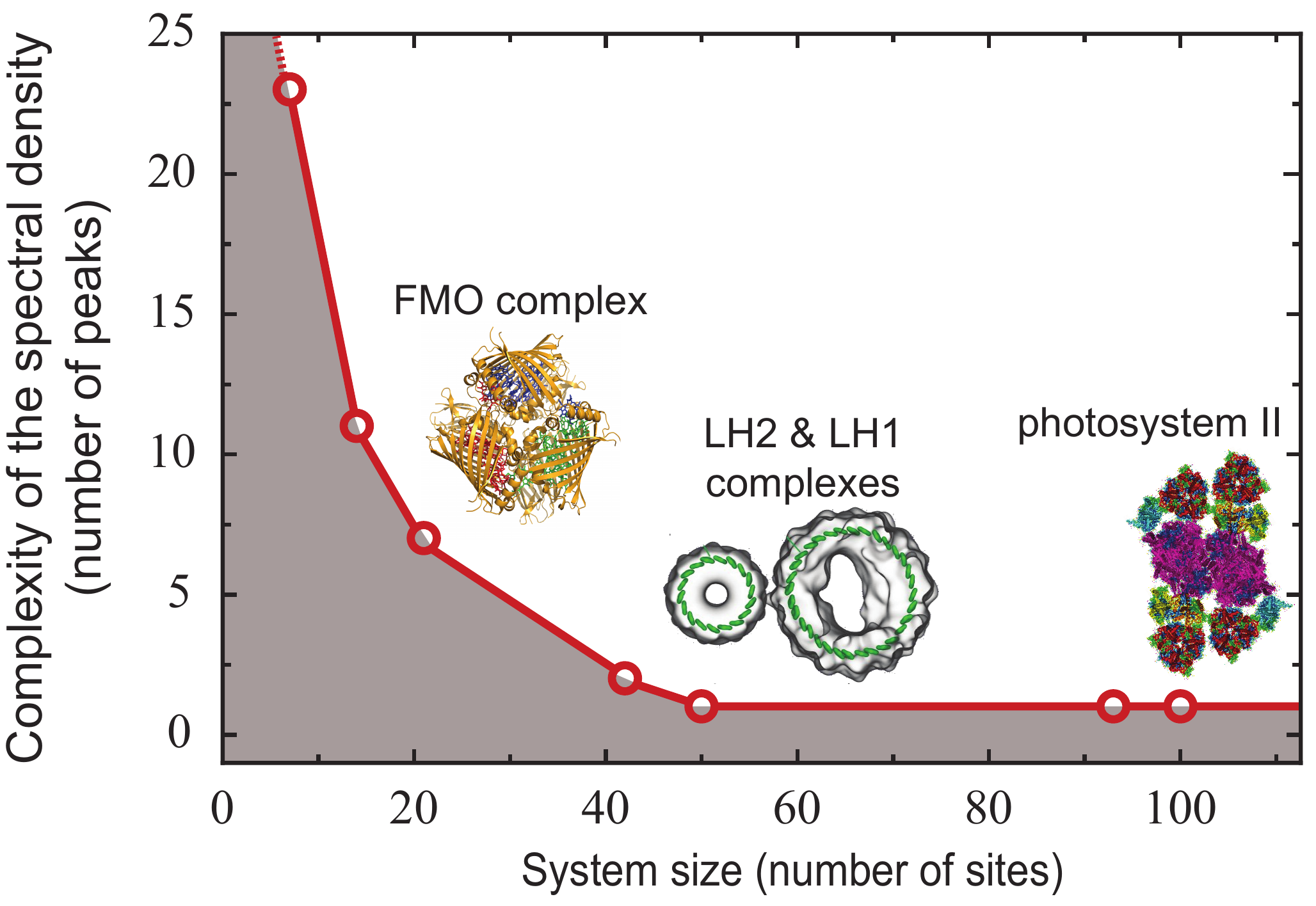}}
\caption{
%
The grey area shows the estimated treatable system sizes for the simulation of Frenkel exciton Hamiltonians using current classical supercomputing resources. 
There is a trade off between the complexity of the spectral density and the system size that denotes the classically-feasible area. 
Three photosynthetic systems are shown: The Fenna-Mathews Olson (FMO) complex of Green-Sulfur Bacteria, 
the Light Harvesting I and II complexes of Purple Bacteria and Photosystem II of higher plants. 
The simulation has been performed using the hierarchical equations of motion (HEOM) approach on 64 AMD Opteron 
cores employing a total of 250 GB of RAM.
}\label{plot}
\end{center}
\end{figure}

%
There is a growing interest in understanding the dynamics of open quantum systems,
particularly, when a particle is coupled to a vibrational environment. 
Such situations arise in quantum chemistry and condensed matter physics, for example, in  
photosynthetic complexes or molecular aggregates.
%
Thus, a detailed study of the dynamics of electron-phonon interaction becomes desirable.
Although many analytical and numerical methods have been applied to this problem
\cite{alexandrov1995,freericks1993,ciuchi1997,bonca1999,romero1999,hoffmann2002,barisic2004,
spencer2005,mona2006,goodvin2006,prokofev2008},
their applicability is often limited by the number of the phonon modes coupled to the electronic states or
to a particular investigation (e.g. low-lying excited polaron) and a specific parameter regime.
The resources required for most of the classical computational
methods increase exponentially with the number of particles in the simulation
and it is challenging to simulate the dynamics of open quantum systems on
conventional computers, even using modern parallel processing units \cite{Christoph2011,Christoph2012,Christoph2014}.
The situation becomes even much more challenging for complex open quantum systems with structured environments.
As yet, only small model systems have been studied theoretically with crude approximations to the system-bath 
dynamics, see for example~\cite{fujita2012,fujita2013}. 
Numerically exact solution can be obtained for only small systems ($<$ 20 sites) 
with restrictions on the bath modes~\cite{Ishizaki2009,Ishizaki2009-2,Christoph2011,Christoph2014,Strunz1999,Alex2014}.
%
In Figure~\ref{plot}, we estimate the upper limit for simulating such complex open quantum systems
with current computational resources on conventional computers.
The horizontal axis indicates the system size (number of the particles or sites) that can be simulated 
while the vertical axis indicates number of the peaks in the spectral density that could be considered in 
this simulation, see Ref.~\cite{Christoph2014} for more computational details. 
Note that ``peaks'' here refers to  Drude-Lorentz peaks in the spectral density~\cite{Christoph2014} which 
should not be confused with the number of phonon modes in the Hamiltonian.
Each of these peaks in the spectral density may include several phonon modes. 
%

In this work, we propose analog quantum devices \cite{feynman,nori,mostame2011} to mimic the 
dynamics of complex open quantum systems and demonstrate that they can be constructed using 
present-day technology of superconducting circuits and outperform current classical computational methods.
With such quantum simulators, one can perform more extensive investigation including exciton transport,
spectral density, absorption spectra
as well as  wide range of  parameters and thereby a more detailed understanding of the problems.
Furthermore, our proposed quantum simulators occupy a wide region in the plot shown in Figure~\ref{plot}.
Similar ideas for simulating Holstein polarons based on polar molecules trapped in an optical lattice \cite{herrera2011,herrera2013},  
Rydberg states of cold atoms and ions \cite{hague2012}, trapped ions \cite{cirac2012,solano2012}
and  superconducting circuit quantum electrodynamics (QED) \cite{mei2013,Stojanovic2014} have been pursued earlier. 
However, the main focus of this paper is emulating the dynamics of multiple-mode Holstein models at 
finite temperature -- with application in open quantum systems with complex environments --
which has not been addressed in any of the above mentioned references.  
Moreover, it is worthwhile to study an alternative set-up, since different experimental
realizations carry distinct advantages and drawbacks.
%

%
\textbf{\em Standard Holstein model}.
We first focus on simulating an electron-phonon model which describes the interaction of 
a single electron on a 1D finite lattice with one vibrational mode 
per lattice site, namely the Holstein model:
\bea
\label{HolsteinHamil}
H_{\rm Hol} = H_{\mathrm{el}}+H_{\mathrm{ph}}+H_{\mathrm{el-ph}}\, .
\ea
The first term of the above Hamiltonian (electronic term) is given by
\mbox{$ H_{\mathrm{el}}=\sum_{n=1}^{N-1} V_n \left(a_n^\dagger a_{n+1} + a_{n+1}^\dagger a_n \right)$}
with $V_n$ being the strength of the nearest-neighbor couplings,  
$a_n^\dagger \,\, (a_n)$ being the creation (annihilation) operators of the electron and $N$ being the number of sites.
The phonon Hamiltonian is \mbox{$H_{\mathrm{ph}} = \sum_{n=1}^N \hbar \omega_n b_n^\dagger b_n$},
with $\omega_n$ being the frequency of the phonon mode coupled to the $n$-th lattice site
and $b_n^\dagger\, \, (b_n)$ being the creation (annihilation) operators of the phonon.
The last term in Eq.~(\ref{HolsteinHamil}) describes the electron-phonon coupling
$H_{\mathrm{el-ph}}=\sum_{n=1}^N \kappa_n \, a_n^\dagger a_n \left(b_n^\dagger + b_n \right)$
with $\kappa_n$ being the coupling strength between the electron and phonon at lattice site $n$.
Using the Jordan-Wigner transformation, the Hamiltonian $H_{\rm Hol}$ can be rewritten in 
terms of the Pauli {\boldmath$\sigma$} operators,
\bea
\label{HolsteinHami2}
H_{\rm Hol} &=& \frac{1}{2} \sum_{n=1}^{N-1} 
\,V_n\, \left(\sigma_x^n \sigma_x^{n+1} +  \sigma_y^n \sigma_y^{n+1} \right)+
\nn
&&
+ \sum_{n=1}^N \left[ \kappa_n\, \sigma_z^n \left(b_n^\dagger + b_n \right) 
+ \hbar \omega_n b_n^\dagger b_n \right] \, .
\ea

\begin{figure}[t]
\begin{center}
\resizebox{\linewidth}{!}{\includegraphics{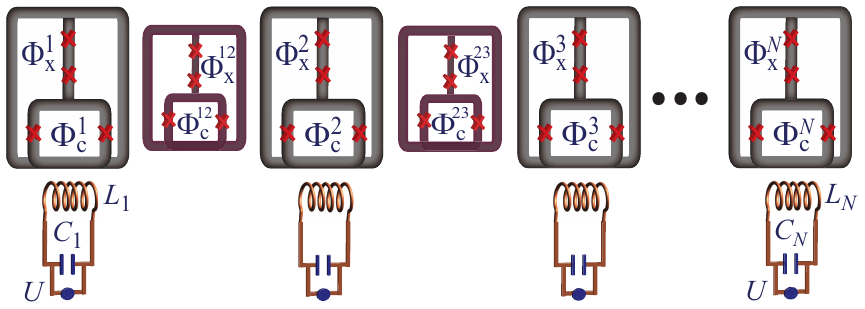}}
\caption{Superconducting quantum circuit diagram of the proposed quantum simulator for the Holstein model.
The qubit states are encoded in the quantized circulating current 
of the qubit loop.
The red crosses denote Josephson junctions.
The gradiometric flux qubits are coupled with a tunable $\sigma_z \sigma_z$-coupling.
Each of the qubits is independently coupled to a quantum LC-oscillator
to simulate the vibrational environment.
}\label{simu1}
\end{center}
\end{figure}

In order to reproduce the quantum dynamics of the open system given by the above Hamiltonian,
let us consider a chain of $N$ gradiometric flux qubits 
\cite{mostame2011,paauw2009} with tunable $\sigma_z \sigma_z$-couplings 
\cite{lloyd2007,ashhab2008} and a single LC-oscillator coupled to each qubit, 
as shown in Figure~\ref{simu1}.
The Hamiltonian of a single flux qubit in the bare basis, the quantum states with magnetic 
flux pointing up $\ket{\uparrow}$ and down $\ket{\downarrow}$, is given by 
\mbox{$H_{\rm q}^i=  \, (\mathscr{E}_i \,\sigma_z^i + \Delta_i \,\sigma_x^i)/2$} \cite{clarke}, where $\mathscr{E}_i$ 
is the energy bias between $\ket{\uparrow}$ and $\ket{\downarrow}$, $\Delta_i$ is 
the tunnel splitting between the two states and $i$ labels the position of the qubit in the chain.
Note that $\mathscr{E}_i$ can be tuned to zero to neglect the term \mbox{$\mathscr{E}_i\sigma_{z}^i$} 
and therefore be at the optimal operating point \cite{bertet2005} of the flux qubit, which is the most 
common case in current experiments.
The coupling between two nearest-neighbor qubits in the bare basis is given by 
\mbox{$ H_{\rm coup}^i=  g_i(\Delta^c_{ii+1}) \, \sigma_z^i\, \sigma_z^{i+1}$},
where $\Delta^c_{ii+1}$ is the (tunable) tunnel splitting of the coupler qubit (smaller
qubits in Figure~\ref{simu1}, see Ref.~\cite{mostame2011} for more details).
The coupling of a quantum LC-oscillator to the smaller loop of a flux qubit, as shown in 
Figure~\ref{simu1}, is given by
\mbox{$H_{\rm q-osc}^i=\eta_i \, \sigma_x^i \left(c^\dagger_i + c_i \right) $} with 
$c_i^\dagger \, (c_i)$ being the creation (annihilation) operator of the oscillator coupled 
to the $i$-th qubit and $\eta_i$ being the coupling strength.
Finally, the Hamiltonian of a single oscillator is
\mbox{$H_{\rm osc}^i=\hbar \omega'_i\,  c_i^\dagger c_i$} with
$\omega'_i$ being the transition frequency of the oscillator.
Rewriting the above Hamiltonians in the energy eigenbasis of the qubit
\mbox{$\ket{\pm}=\left(\ket{\downarrow}\pm\ket{\uparrow}\right)/\sqrt{2}$}
converts the operators \mbox{$\sigma_x^i\, \to \, \sigma_z^i$} and 
\mbox{$\sigma_z^i\, \sigma_z^{i+1} \, \to \, \sigma_x^i\, \sigma_x^{i+1} \approx
\left( \sigma_x^i \sigma_x^{i+1} + \sigma_y^i \sigma_y^{i+1} \right)/2$}
in the rotating wave approximation (neglecting strongly off-resonant couplings).
Then the total Hamiltonian of the superconducting circuit proposed to 
emulate the dynamics of the Holstein model 
\mbox{$H_{\rm sim}=\sum_{i=1}^N \left\{H_{\rm q}^i+H_{\rm coup}^i+H_{\rm q-osc}^i+H_{\rm osc}^i\right\}$} 
in the new basis is given by 
%
\bea
\label{SimuHami1}
H_{\rm sim} &\approx& \frac{1}{2} \sum_{i=1}^{N-1}  
g_i(\Delta^c_{ii+1}) \left[ \sigma_x^i\, \sigma_x^{i+1}+  \sigma_y^i\, \sigma_y^{i+1}\right]
\nn
&&
+ \sum_{i=1}^N \left\{ \frac{\Delta_i}{2} \,\sigma_z^i + \eta_i \, \sigma_z^i \left(c_i^\dagger + c_i \right) + \hbar\omega'_i\,c_i^\dagger c_i \right\}\, . \nn
\ea
Here we can assume identical flux qubits, similar to the most common case considered in the literature of the Holstein model.
Then the term $\sum_{i=1}^N \Delta_i/2 \,\sigma_z^i$  in the above Hamiltonian leads to a 
global phase.
Comparison of the Hamiltonians~(\ref{HolsteinHami2}) and (\ref{SimuHami1}), demonstrates that
a chain of coupled flux qubits with a single quantum LC-oscillator coupled to each qubit can simulate 
the same dynamics of the Holstein model with $g_i(\Delta^c_{ii+1}), \, \eta_i, \, \omega'_i$ 
corresponding to $V_n, \, \kappa_n, \, \omega_n $, respectively.
Interestingly, for superconducting flux qubits, the couplings $g_i(\Delta^c_{ii+1})$ and $\eta_i$ are tunable.
The implementable range of $g_i(\Delta^c_{ii+1})$ is in the 
range of approximately zero to 1~[GHz] \cite{niskanen2007}.
$\eta_i$ can be in the $<\,$10~[GHz] range depending on the frequency of the resonator. 
The tunability and wide implementable range of these parameters makes it possible to study different 
parameter regimes of interest (strong coupling $\eta_i \gg g_i$, weak coupling $\eta_i \ll g_i$ and
intermediate $\eta_i \sim g_i$  regimes) using the proposed quantum simulator.
%
%

Preparation of the qubits in their ground state is straightforward: 
One needs to allow them to relax as close as possible to their ground state by cryostatic cooling.
Subsequently, the qubits can be initialized by flux control in the appropriate initial states for the simulation.	
The excitation of a qubit is undemanding to achieve with the 
application of a resonant microwave excitation ($\pi$-pulse) carried by a microwave 
line which is connected to the respective qubit.
This technique has been used extensively, e.g., for the observation of 
Rabi oscillations in a flux qubit \cite{clarke,chiorescu}.
After some evolution time the populations  of the qubit states are measured.
%

\textbf{\em Temperature}.\quad  
Note that the standard Holstein model discussed above does not contain temperature, however, the superconducting 
circuit~(as a real physical system) is at finite temperature $T_{\rm sim}$. 
%
%
Currently, a superconducting circuit can be refrigerated down to a very low temperature, around 10~[mK]$\,\approx\,$0.2~[GHz],
and the flux qubits can be even cooled down far below 10~[mK] using active microwave cooling~\cite{Valenzuela2006}.
%
%
Although the quantum simulator being at finite temperature seems to be a disadvantage, 
we will see in the following that the easy tunability over $T_{\rm sim}$ 
allows one to investigate the physically relevant case of a finite temperature Holstein model over a wide range of temperatures.
%
%
To this end, we will generalize the Standard Holstein Hamiltonian.

\begin{figure}[t]
\begin{center}
\resizebox{0.9\linewidth}{!}{\includegraphics{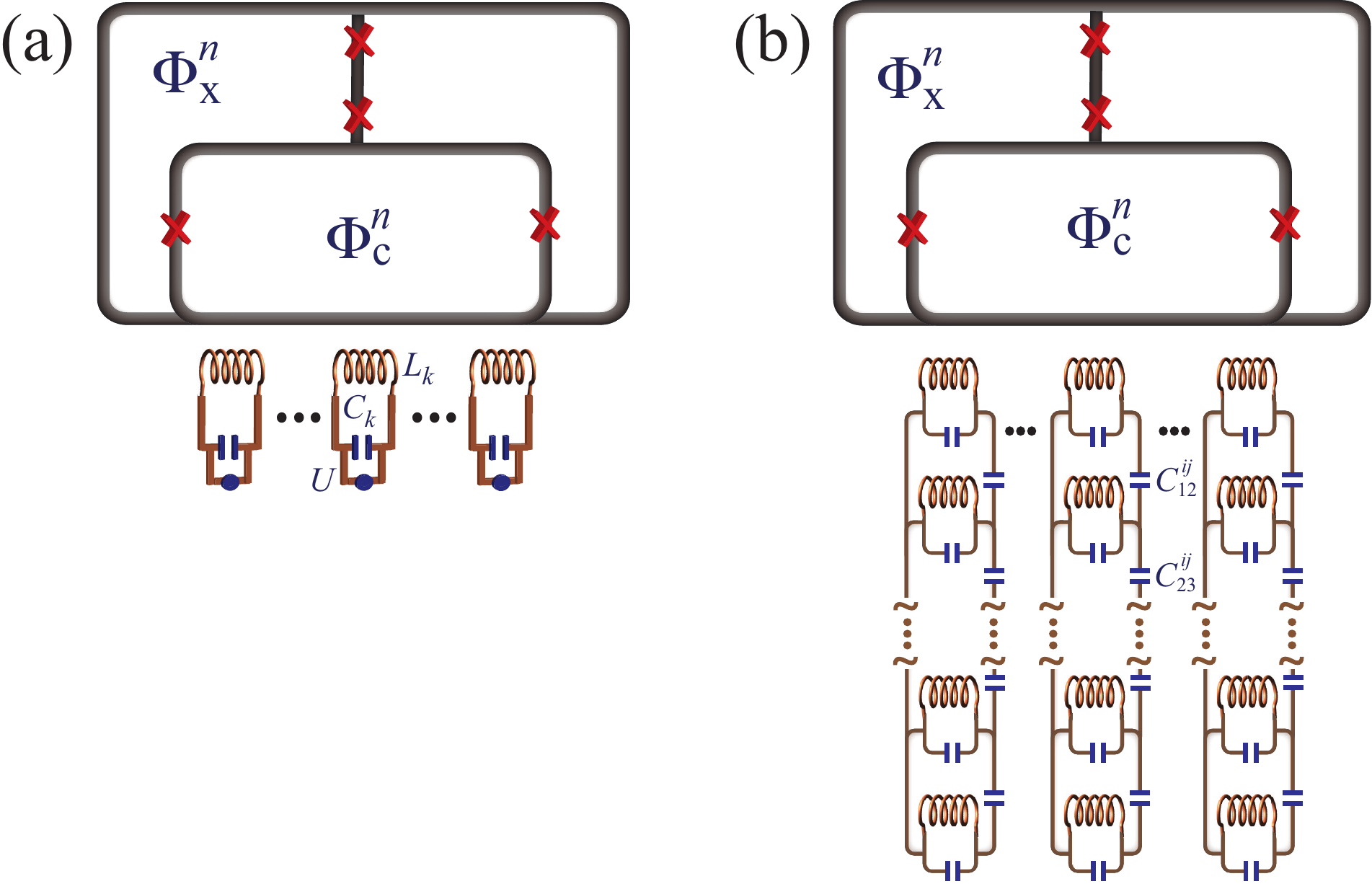}}
\caption{Representation of a single flux qubit coupled to quantum LC-oscillators. 
(a) Many single resonators are directly coupled to the qubit.
(b) Using a linear-algebraic bath transformation~\cite{Joon2014}, the set of independent resonators (directly coupled to 
the qubit) are transformed into a set of weakly-coupled multiple parallel chain of resonators.}
\label{fig3}
\end{center}
\end{figure}

\textbf{\em Generalized Holstein model}.\quad  
The Hamiltonian of multi-mode Holstein model is given by
\bea
\label{aggregatesHami}
&& H_{\rm gen} = \frac{1}{2} \sum^N_{\substack{n=1 \\ n\ne m}}
\sum_{m=1}^N \,J_{nm}\, \left(\sigma_x^n \sigma_x^{m} +  \sigma_y^n \sigma_y^{m} \right)
\nn
&&
+ \sum_{n=1}^N \left\{ \sum_k \left[ \kappa_{nk}\, \sigma_z^n \left({b^\dagger_{nk}} + b_{nk} \right)
+ \hbar \omega_{nk} \, {b^\dagger_{nk}} b_{nk} \right]
+ \mathcal{C}_n \right\}\, ,
\nn
\ea
where $k$ labels the vibrational modes coupled to the site $n$ with frequency
$\omega_{nk}$, $\kappa_{nk}=\hbar \omega_{nk} \sqrt{R_{nk}}$ is the coupling of the electronic excitation of the site $n$
to the vibrational mode $k$ with $R_{nk}$ being the dimensionless Huang-Rhys factor (electron-phonon coupling constant)~\cite{roden2011}, 
and constant \mbox{$\mathcal{C}_n= \epsilon_n +\sum_k \hbar \omega_{nk}R_{nk} + D_n $} with
$\epsilon_n$ being the electronic transition energy,  and $D_n$ being the gas-to-crystal shift 
of the transition energy due to nonresonant forces \cite{roden2011,spano2010}.
Since a constant energy offset does not alter the dynamics, we ignore $\mathcal{C}_n$ 
in our approach.
Now each site couples to a set of oscillators with frequencies $\omega_{nk}$ and corresponding couplings $\kappa_{nk}$.
We have also generalized interactions $J_{nm}$ between arbitrary sites.
The dynamics of a multiple-mode Holstein model can be reproduced by a similar superconducting circuit 
shown in Figure~\ref{simu1} with additional quantum LC-oscillators coupled to each flux qubit, see Figure~\ref{fig3}~(a). 
%
%

The experimental implementation of such a quantum simulator can face challenges due to the current 
constraints in the realizable superconducting circuits. 
The number of quantum LC-oscillators, that are directly coupled to a qubit is limited by the physical size of the
superconducting qubits. 
Moreover, the coupling strength of the qubit to the quantum oscillator is limited and should not exceed a 
certain percentage of the frequency of the oscillator~\cite{Jamie2012}. 
A simple numerical formula for the coupling $\kappa_{nk}$ is given by,
\bea
\label{QRCoupling}
&& \frac{\kappa_{nk}}{\hbar \, \omega_{nk}} \, = \, \sqrt{R_{nk}} \, = \,
\nn
&& 
= \, \frac{5.48 \,\beta_{nk} \, I^{nk}_p}{50\,[{\rm nA}]} 
\left(\frac{Z^{nk}_r}{100\,[{\rm Ohms}]}\right)^{1/2}
\left(\frac{\omega_{nk}}{2\pi\, [{\rm GHz}]}\right)^{-1}.
\ea
Here $\beta_{nk}$ is the inductive division ratio (flux of the $k$-th oscillator coupled to the $n$-th qubit is $\beta_{nk}$ times of the qubit flux).
This parameter needs to be far below 1 to avoid hybridizing the qubit with the resonator.
$Z^{nk}_r$ is the oscillator impedance and has to be well below the impedance of free space (not much higher than 100~[Ohms]),
in order to maintain high quality factors for the resonators.
$I^{nk}_p$ is the effective persistent current of the DC SQUID loop, which is the linear slope of the qubit energy splitting with 
respect to DC SQUID flux.
In principle, $I^{nk}_p$ can be made large, but this also can cause linear flux sensitivity of the qubit energy. 

These challenges can be addressed and resolved by a linear-algebraic bath transformation~\cite{Joon2014}
that we have proposed recently.
Based on a simple linear algebraic approach, the set of independent LC oscillators directly coupled to 
a qubit, Figure~\ref{fig3}~(a), can be transformed into a set of weakly-coupled multiple parallel chain of 
oscillators, see Figure~\ref{fig3}~(b). 
This transformation can dramatically reduce the number of the oscillators that are directly coupled to the qubit as well as
the coupling strength of the quantum oscillators to the qubit. 
To specify the number of the required resonators and their parameters and to feature outrunning classical 
algorithms with our proposed approach, as an example, here we study the feasibility and provide an outlook for the
emulation of the dynamics of the chlorosome light-harvesting antenna from green sulfur bacteria.

\begin{figure}[t]
\begin{center}
\resizebox{\linewidth}{!}{\includegraphics{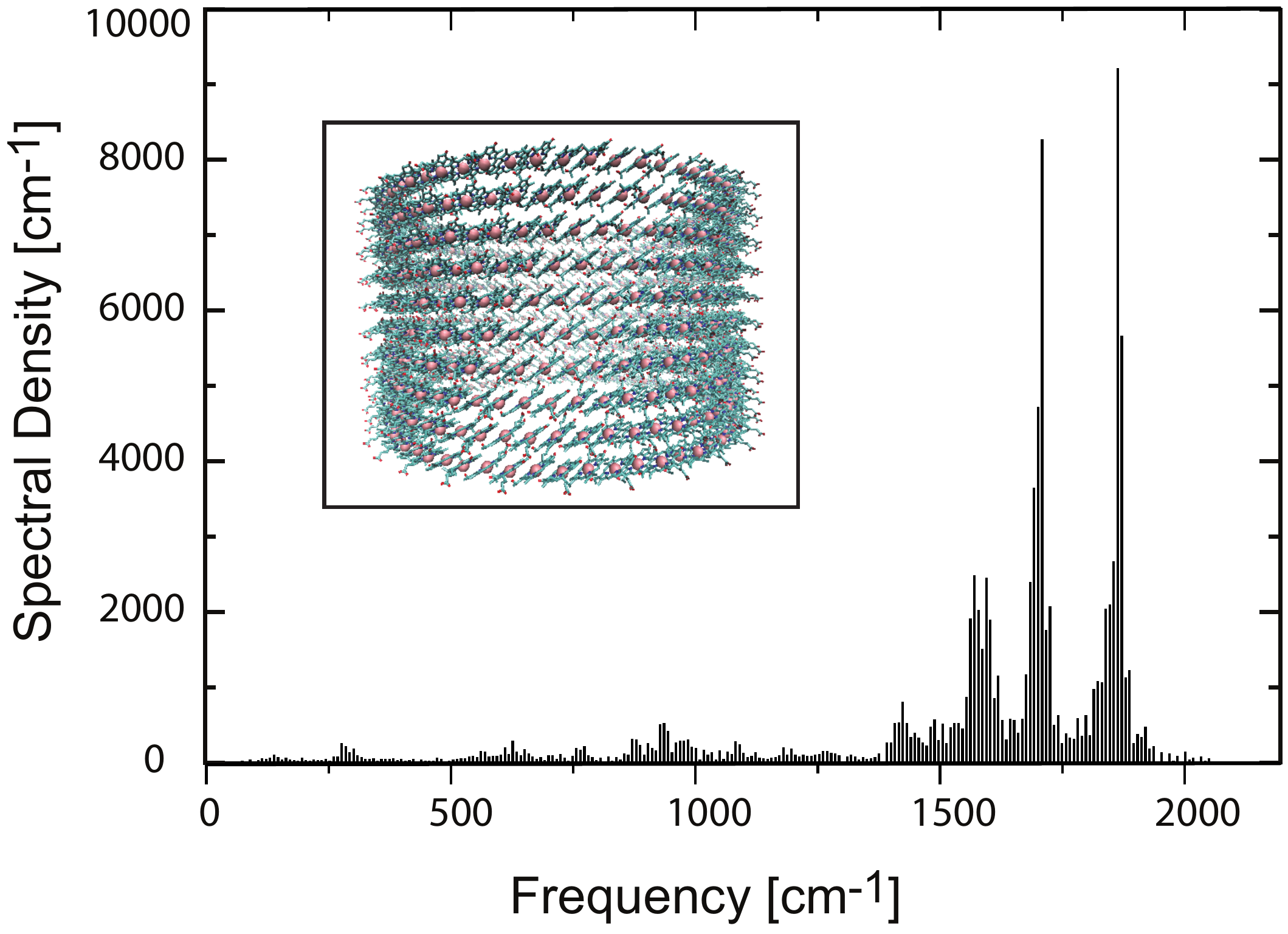}}
\caption{Spectral density of the electron-phonon coupling of bacteriochlorophyll molecules in the chlorosome 
antenna of green-sulfur bacteria with 253 phonon modes. The spectral density 
of the phonon bath was obtained from a quantum mechanics/molecular mechanics (QM/MM) 
simulation with time-dependent density function theory in Ref.~\cite{fujita2012}.  
An experimentally-resolved \cite{Ganapathy2009} structure of the chlorosome is shown in the inset.
}\label{sdchlorosome}
\end{center}
\end{figure}

\textbf{\em Chlorosome light-harvesting antennae}. \quad
The green sulfur bacteria lives in a deep sea where only a few hundred photons per second arrive at a bacterium~\cite{manske2005}. 
Therefore, they should be able to transfer the photon energy efficiently, rapidly and robustly to the reaction center 
to generate the electro-chemical potential energy gradient and exploit it in the photosynthetic metabolic cycle.
%
Compared with other light-harvesting species, the chlorosome has the unique feature that 
it is composed of 200--250 thousands bacteriochlorophyll molecules organized into supramolecular 
assembly~\cite{Oostergetel2010,Ganapathy2009}.
How the quantum dynamics helps the excitation energy transfer within this giant molecular aggregate is an 
interesting question and has attracted many research groups, 
see the references cited in Refs.~\cite{Oostergetel2010,fujita2012,fujita2013}. 
%
%
The dynamics of the chlorosome can be given by the multi-mode Holstein model in Eq.(\ref{aggregatesHami}). 
%
The structure model has been proposed from experiments~\cite{Ganapathy2009} and studied theoretically by some of the 
authors~\cite{fujita2012,fujita2013} with a crude stochastic quantum propagating model and its spectral density is demonstrated
in Figure~\ref{sdchlorosome}.

As discussed above, the dynamics of this system can be emulated by a chain of superconducting 
qubits and 253 quantum LC-oscillators coupled to each qubit. 
The size of each flux qubit is around few ten of microns and there is no enough physical space to couple it directly to 253 resonators.
However, we can reduce number of the resonators that are directly coupled to the qubit by using the linear-algebraic bath 
transformation~\cite{Joon2014}. 
This transformation mixes resonator modes with different frequencies to distributes 253 modes to, for example, 
a set of 6 parallel chains of quantum resonators,~Figure~\ref{fig3}~(b), with each of the chains having the most of 43 coupled oscillators. 
In addition to reducing number of the resonators that are directly coupled to the qubit, 
this mapping will also reduce the coupling strength of the qubit to the primary oscillator 
modes (the first oscillators in the chains that are directly coupled to the qubit).

Note that the parameters of the superconducting simulator are temperature dependent.
To account for finite temperature,
we first transform the spectral density given in Figure~\ref{sdchlorosome}
using \mbox{$C(\omega,T)=\left\{1+\coth\left[\hbar\omega/(2 k_BT) \right]\right\} J^{\rm A}(\omega)$},
where the subscript ``A'' denotes the antisymmetric spectral density
$J^{\rm A}(\omega)= J(\omega)$  if $\omega \ge 0\, $; and
$J^{\rm A}(\omega)= - J(-\omega)$ if $\omega < 0\,$, see~\cite{mostame2011} for more details. 
Since the chlorosome is at room temperature, $T_{\rm ch}$=300~[K], and the superconducting circuit can be considered at 
$T_{\rm sim}$=10~[mK], then all the parameters of the quantum simulator need to be rescaled accordingly; 
\mbox{$\mathcal{A}_{\rm sim}=\left(T_{\rm sim}/T_{\rm ch}\right) \mathcal{A}_{\rm ch}$}, with 
$\mathcal{A}_{\rm sim}$ and $\mathcal{A}_{\rm ch}$ indicating any parameter of the quantum simulator and chlorosome, respectively. 
Then after rescaling, we perform the linear-algebraic bath transformation.
With this procedure, the coupling strengths between the qubit and the oscillators that are directly
coupled to the qubit,~Figure~\ref{fig3}~(b), need to be around \mbox{150 -- 210~[MHz]}.
The coupling between the oscillators in the chains are around \mbox{100 -- 560~[MHz]}, the required frequencies for 
the quantum oscillators are around  \mbox{1.4 -- 1.6~[GHz]}.
The resonators here need to have high quality factors.  

\textbf{\em Conclusion and Outlook}.\quad 
We have shown that it is appealing to simulate the dynamics of open quantum systems with complex environments and structured spectral densities 
(such as, the chlorosome or the examples given in Figure~\ref{plot}) by using a chain of few tens of coherent qubits.
In our previous work~\cite{mostame2011}, we presented a detailed study on simulating the dynamics of Fenna-Matthews-Olson photosynthetic complex as an example of complex open quantum systems.
The main focus of current manuscript has been to address the limit that analog quantum simulators based on superconducting circuits 
with precisely engineered quantum environment may outperform exact classical computational approaches, 
such as the HEOM approach, allowing us to study non-Markovian effects.
Furthermore, here we have discussed the simulation of standard, as well as, generalized Holstein model at finite temperature which
has many applications in molecular aggregates, polymers, and superconductivity.
%
%
Using the linear-algebraic bath transformation, we will be able to simulate dynamics of complex open 
quantum systems with thousands of phonon modes. 
Such a simulation is as exact as numerical approaches such as HEOM and definitely out of reach of any currently available 
computational device.

\textbf{\em Acknowledgments}.\quad
We acknowledge DTRA Grant No.~HDTRA1-10-1-0046, AFOSR UCSD Grant No.~sponsor Award Number: 10323836-SUB,   
Prime award number FA9550-12-1-0046 and Harvard FAS RC team for Odyssey computer resources.

$^*$\,{\footnotesize\sf aspuru@chemistry.harvard.edu}
$^\dagger$\,{\footnotesize\sf eisfeld@mpipks-dresden.mpg.de}


\end{document}